\documentclass[pdflatex,sn-nature]{sn-jnl}

\usepackage{graphicx}
\usepackage{amsmath,amssymb,amsfonts}
\usepackage{booktabs}
\usepackage{hyperref}

\begin{document}

\title{Empirical confirmation of bosonic wealth statistics in Bitcoin UTXOs}

\author[1]{\fnm{Chanhee} \sur{Park}}

\author*[2]{\fnm{Claudio J.} \sur{Tessone}}\email{claudio.tessone@uzh.ch}

\author[3]{\fnm{Yu} \sur{Zhang}}

\author*[1]{\fnm{Jeong-Hyuck} \sur{Park}}\email{park@sogang.ac.kr}

\affil*[1]{\orgdiv{Department of Physics}, \orgname{Sogang University}, \orgaddress{\city{Seoul}, \postcode{04107}, \country{Korea}}}

\affil[2]{\orgdiv{Blockchain and Distributed Ledger Technologies, IfI, UZH Blockchain Center}, \orgname{University of Zurich}, \orgaddress{\city{Zurich}, \postcode{8050}, \country{Switzerland}}}

\affil[3]{\orgdiv{Blockchain and Distributed Ledger Technologies, IfI}, \orgname{University of Zurich}, \orgaddress{\city{Zurich}, \postcode{8050}, \country{Switzerland}}}

\abstract{Digitalisation transforms money from distinguishable physical objects into fungible informational units. A recent theoretical framework predicts that such indistinguishable wealth obeys bosonic occupancy statistics, leading to geometric ownership distributions and enhanced inequality. Using Bitcoin blockchain data, we test this prediction on 63 UTXO denominations across 72 monthly snapshots (2018--2023). A one-parameter geometric model describes the ownership distributions, reproducing both mean holdings and their temporal evolution; Jensen--Shannon divergence values lie below $0.08$ in $99.74\%$ of cases. The inferred inverse-temperature parameter satisfies the analytic mean--temperature relation to better than $0.1\%$ in every sample---a self-consistency test that two-parameter alternatives cannot pass---and remains within a narrow band across eight orders of magnitude in denomination and over six years. Bitcoin UTXO ownership statistics are therefore consistent with bosonic occupancy laws, suggesting that the informational nature of electronic money may act as a structural driver of inequality in digital economies.}


\maketitle

\section{Introduction}\label{sec:intro}

The digitalisation of the global economy is transforming the fundamental nature of money. Physical cash, composed of distinguishable objects such as coins and banknotes, is increasingly being replaced by abstract informational units recorded on digital ledgers. This raises a fundamental question: does the ontological form of wealth influence its statistical distribution?

Following concepts from statistical physics developed in early econophysics work~\cite{Dragulescu2000,Yakovenko2009}, wealth distributions can be analysed within a statistical mechanics framework. This approach complements behavioural and institutional explanations by emphasising that the intrinsic properties of wealth units shape aggregate distributions.

Traditional economic models explain wealth inequality in terms of agent behaviour, exchange rules, or institutional structures. By contrast, the present approach emphasises a more fundamental distinction between distinguishable and identical units, which exhibit qualitatively different statistical behaviours.

From this perspective, recent theoretical work~\cite{KimPark2023} suggests that the intrinsic indistinguishability of digital assets may act as a structural driver of wealth concentration. In particular, identical units are proposed to obey \textit{bosonic statistics}, leading to qualitatively different ownership distributions. Physical cash systems---composed of distinguishable units---are expected to yield Poisson-type ownership distributions characterised by a mode near the mean, whereas indistinguishable digital wealth follows a geometric distribution with a heavier tail that inherently favours accumulation among large holders.

With the rise of cryptocurrencies, new opportunities have emerged to perform large-scale observation of statistical patterns. These blockchain-based systems act as public, open ledgers where all transactions are recorded in a tamper-evident manner. Thus, Bitcoin~\cite{Bitcoin} provides a well-suited system to test this hypothesis empirically. The blockchain records value transfers and the evolving state of unspent transaction outputs (UTXOs) in a single consistent ledger. Unlike traditional banking data, which is proprietary and aggregated, Bitcoin allows for a transparent, denomination-resolved analysis of ownership distributions~\cite{Meiklejohn}. Specifically, by conditioning on fixed UTXO face values, one can probe how discrete units of value are packaged and held across addresses. If the theoretical framework holds, the occupancy distribution of these identical digital units should exhibit the signature of bosonic statistics, regardless of the specific denomination or temporal market conditions.

Despite this intuition, and notwithstanding related work on the Bitcoin transaction network~\cite{Kondor2014}, systematic empirical validation using large-scale blockchain data has been lacking.

We present a systematic empirical test of the bosonic wealth hypothesis using Bitcoin blockchain data spanning January 2018 to December 2023. Analysing 63 representative UTXO denominations across 72 monthly snapshots, we perform a one-parameter Bayesian inference of the ownership distributions. We find that for every denomination and throughout the entire period, the empirical distributions are accurately described by geometric laws characterised by a single inverse-temperature parameter. Information-theoretic comparison using the Jensen--Shannon divergence confirms the robustness of this agreement, with 99.74\% of cases exhibiting divergence values below 0.08. The inferred parameters consistently reproduce both the empirical mean holdings and their temporal evolution, demonstrating that the theory captures not only the static form of the distribution but also its dynamic behaviour.

To the best of our knowledge, this constitutes the first systematic, quantitative test of bosonic wealth statistics in a real-world monetary system. Beyond cryptocurrencies, the results suggest that the electronic form of money itself---through intrinsic indistinguishability---may act as a structural driver of wealth concentration. This informational indistinguishability implies that, unlike Poisson-type distributions characteristic of traditional tangible currencies, geometric distributions place greater probability mass at low holdings while supporting a long exponential tail, inherently favouring accumulation among large holders. We thus propose that the digitalisation of wealth provides a fundamental statistical mechanism for inequality, independent of preferential attachment, strategic behaviour, or institutional asymmetry. In the following sections, we detail the statistical mechanics perspective, present the empirical verification using Bitcoin data, and discuss the implications for digital economies and wealth distribution.

\section{Results}\label{sec:results}

\subsection{Statistical mechanics of indistinguishable wealth}\label{sec:theory}

To address quantitatively whether the form of wealth shapes its statistical distribution, we begin from the combinatorial origin of the two statistical families. Borrowing an analogy from particle physics, wealth can be classified into two fundamental categories---\textbf{distinguishable} and \textbf{identical}---each following distinct statistical laws, as if they were different types of particles. Distinguishable wealth, such as coins or banknotes, behaves like classical distinguishable particles and yields a \textbf{Poisson distribution}. In contrast, identical wealth, such as bank deposits or cryptocurrencies, behaves analogously to a bosonic system, following a \textbf{geometric distribution} when described in terms of holders. In statistical mechanics, the occupancy-number distribution of non-interacting bosons in a single quantum state follows a geometric law~\cite{Huang}, which motivates the statistical analogy used here.

Let $k=0,1,2,\ldots$ denote the discrete, dimensionless number of units (e.g., coins, deposits, or UTXOs) held by a single holder, and let $m=\langle k \rangle$ denote the average holding. The distributions then take the form:
\begin{equation}
	P_{\mathrm{Poisson}}(k)
	= e^{-m}\frac{m^{k}}{k!}\,,
	\qquad
	P_{\mathrm{Geom}}(k)
	= \frac{1}{m+1}\!\left(\frac{m}{m+1}\right)^{k}\!.
	\label{twoP}
\end{equation}
Within the present combinatorial framework, distinguishable wealth units yield the Poisson distribution while identical units yield the geometric distribution. The Poisson distribution peaks near the mean value, representing the broad ``middle-class'' structure typical of distinguishable wealth. In contrast, the geometric distribution decreases monotonically and exhibits a much heavier tail, implying enhanced concentration among large holders.

A previous theoretical framework~\cite{KimPark2023} proposed that electronic assets are inherently identical because they lack physical individuality and exist only as digital records. Their indistinguishability arises from their informational nature: what matters is the amount, not the identity of each unit.

The ownership-based distributions in Eq.~(\ref{twoP}) can be derived from a simple combinatorial argument in analogy with statistical mechanics. Let $N$ denote the total number of holders and $M$ the total number of wealth units, so that the mean holding is $m=\langle k\rangle=M/N$. We denote by $n_k$ the number of holders who each possess $k$ units, subject to
\begin{equation}
	\sum_{k=0}^{\infty} n_k = N\,,
	\qquad
	\sum_{k=0}^{\infty} k\,n_k = M\,.
	\label{eq:constraints}
\end{equation}
These two relations correspond respectively to the conservation of the total number of holders and of the total amount of wealth units.

For a given set $\{n_k\}$, the total number of possible microstates, or degeneracy, is denoted by $\Omega$ and factorises into two parts:
\begin{equation}
\Omega = \Upsilon \times \Phi\,.
\end{equation}
The first factor $\Upsilon$ counts the number of ways $N$ distinguishable holders can be grouped according to $\{n_k\}$,
\begin{equation}
\Upsilon = \frac{N!}{\prod_{k} n_k!}\,,
\end{equation}
while $\Phi$ counts the number of ways the $M$ units of wealth can be arranged among the holders. For distinguishable (classical) units, the total number of permutations of $M$ labelled units is $M!$; dividing by $(k!)^{n_k}$ for each group of holders with $k$ units---whose internal ordering is irrelevant to the macrostate---gives
\begin{equation}
	\Phi_{\mathrm{distinguishable}} =
	\frac{M!}{\prod_k (k!)^{n_k}}\,,
	\label{eq:PhiDist}
\end{equation}
whereas for identical (electronic) units, each configuration of $\{n_k\}$ corresponds to a single microstate,
\begin{equation}
	\Phi_{\mathrm{identical}} = 1\,.
	\label{eq:PhiId}
\end{equation}
The distinction between Eqs.~(\ref{eq:PhiDist}) and~(\ref{eq:PhiId}) is the essential difference between distinguishable and identical wealth. The equilibrium ownership distribution corresponds to the most probable macrostate, obtained by maximising $\ln\Omega$ under the constraints (\ref{eq:constraints}). Maximising via Lagrange multipliers, we obtain two distinct forms: the Poisson distribution for distinguishable units and the geometric (Bose--Einstein occupancy) distribution for identical units. With $P(k)=n_{k}/N$, these are precisely the functional forms (\ref{twoP}).

To establish that these statistical laws are not merely formal results but also arise dynamically, we simulate two minimal exchange processes that conserve both the total number of holders and total wealth, transferring a single unit at each step.
\textbf{(i) Random wealth-to-receiver process:} At each step, one unit is selected uniformly at random from the $M$ distinguishable units (equivalently, a holder is chosen with probability proportional to their current holdings), removed from that holder, and assigned to a uniformly chosen receiver among $N$ holders. This random redistribution of distinguishable units yields a stationary distribution converging to the Poisson law.
\textbf{(ii) Random sender-to-receiver process:} At each step, a sender with at least one unit (i.e., $k\geq 1$) transfers one unit to a randomly chosen receiver. This process represents exchanges of identical (bosonic) wealth and produces a stationary distribution converging to the geometric law.

\begin{figure}[ht]
	\centering
	\includegraphics[width=0.5\textwidth]{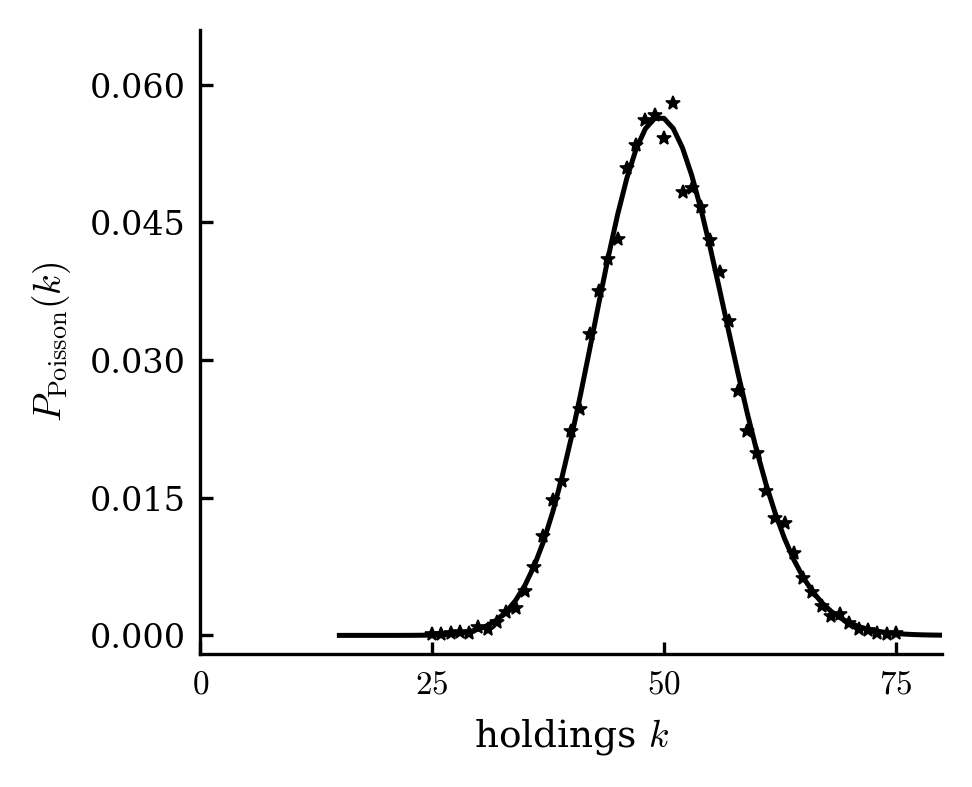}~
	\includegraphics[width=0.5\textwidth]{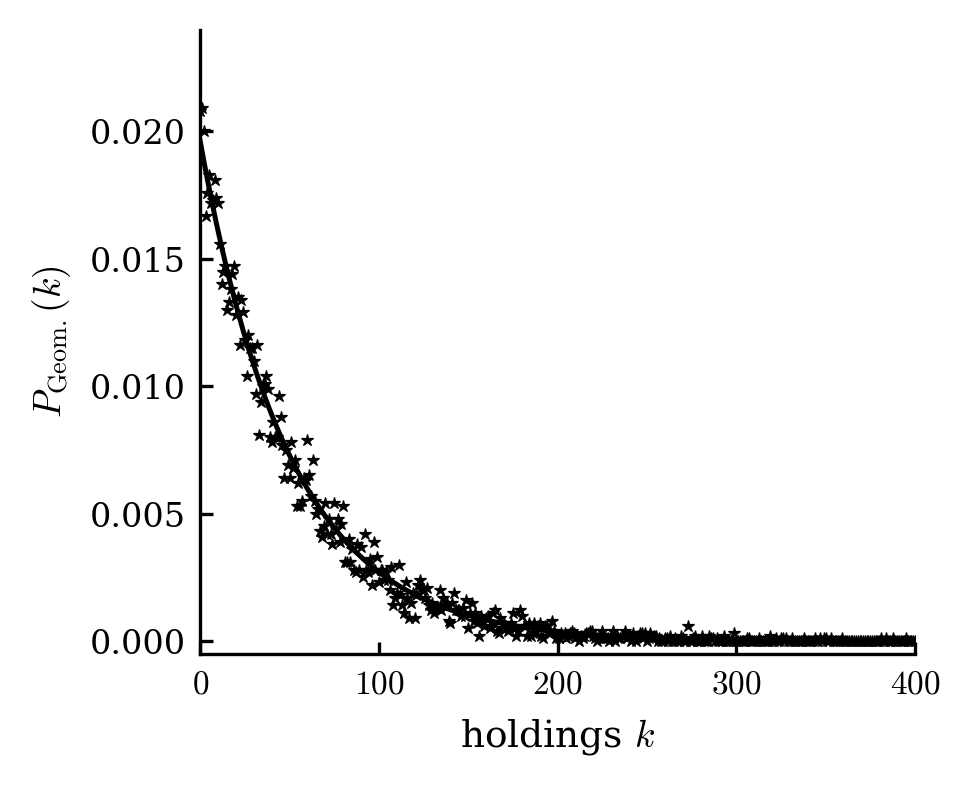}
	\caption{Simulation results for the two minimal exchange models with ${N=10^4}$ and ${M=5\times 10^5}$ after $10^8$ exchange steps. {\bf{Left}}: wealth-to-receiver process (Poisson). {\bf{Right}}: sender-to-receiver process (geometric). Solid lines show the theoretical curves. These simulations serve as a theoretical consistency check and are shown for comparison with the empirical Bitcoin UTXO distributions analysed in the following sections.}
	\label{fig:Simulation}
\end{figure}

Figure~\ref{fig:Simulation} shows representative outcomes of the two simulations. In both cases, the empirical histograms (dots) closely match the theoretical curves (solid lines) of the Poisson and geometric distributions. These results demonstrate that the ownership-based distributions derived from the combinatorial argument also emerge dynamically from minimal stochastic exchange processes.

\subsection{Bitcoin UTXO dataset construction}\label{sec:data}

We analyse Bitcoin mainnet data over 72 monthly snapshots from January 2018 to December 2023 and focus on 63 representative UTXO denominations. The raw data are obtained from a fully synchronised Bitcoin archive node, from which blocks, transactions, and the implied UTXO set are reconstructed using a custom analysis pipeline (BlockSci~\cite{kalodner2020blocksci}). Because the node enforces Bitcoin consensus rules, the reconstructed ledger state is consistent with those rules by construction.

In our analysis, a ``holder'' is an individual Bitcoin address, and we deliberately do not apply address-clustering heuristics in order to avoid introducing model-dependent biases in the empirical distributions. Operationally, an ``address'' is a user-facing encoding that enables the construction of a standard output locking condition (script) such as Pay-to-PubKey-Hash (P2PKH) or Pay-to-Script-Hash (P2SH). For each snapshot, we assign each UTXO to an address if the UTXO's output locking condition corresponds to that address under standard templates supported by our parser, and we then compute denomination-resolved occupancies.

For each snapshot, we study ownership distributions within fixed denominations $i$ (measured in satoshis). These denominations were selected because they are overrepresented face values in the UTXO set over the observation period, rather than being defined through value bins. Practically, these values are common because (i) they align with round-number payment conventions (e.g., $10$, $1{,}000$, $10{,}000$ satoshis) and (ii) they correspond to default or template values embedded in wallet software and payment services, leading to repeated creation and holding of UTXOs at these exact amounts.

Formally, for each denomination $i$ at each monthly snapshot, the fundamental dataset is defined as
\begin{equation}
	\mathcal{D}_{i}=\big\{n^{\star}_{k}\,|\,k=1,2,\ldots,k_{\max,i}^{\mathrm{raw}}\big\}\,.
	\label{dataset}
\end{equation}
Throughout this paper, quantities carrying a star superscript ($\star$) denote empirically observed values extracted from Bitcoin blockchain data; for example, $n_k^{\star}$ denotes the number of addresses controlling exactly $k$ UTXOs with denomination $i$ in a given month. The range of $k$ is finite and excludes zero: $1\le k\le k_{\max,i}^{\mathrm{raw}}$, where the per-snapshot upper bound $k_{\max,i}^{\mathrm{raw}}$ is fixed operationally by the occupancy of the single largest holder of denomination~$i$ in that snapshot.

In our analysis, we exclude dust-level outputs from the denomination set and the corresponding ownership statistics. In Bitcoin Core policy, an output is considered dust if its value is below the relay fee threshold required to spend it~\cite{dust}. This exclusion reduces mechanical artefacts driven by fee policy and change-making behaviour at extremely small values, and it focuses the analysis on denominations with meaningful economic interpretation.

The dataset~(\ref{dataset}) provides the total number of holders and total units,
\begin{equation}
	N_{\star}=\sum_{k=1}^{k_{\max,i}^{\mathrm{raw}}} n^{\star}_{k}\,,\qquad
	M_{\star}=\sum_{k=1}^{k_{\max,i}^{\mathrm{raw}}} k\,n^{\star}_{k}\,,
\end{equation}
from which the raw empirical probability and average holding follow
\begin{equation}
	P_{\star}(k)=\frac{n^{\star}_k}{N_{\star}}\,,\qquad
	m_i^{\star}=\frac{M_{\star}}{N_{\star}}=
	\sum_{k=1}^{k_{\max,i}^{\mathrm{raw}}} k\,P_{\star}(k)\,.
	\label{empiricalm}
\end{equation}
The quantity $m_{i}^{\star}$ defines a monthly time series of average holdings for each UTXO denomination. For the inferential analysis below we work with a smoothed version of the occupancy distribution, denoted $\bar{P}_{\star}(k)$, defined in Methods (Eqs.~(\ref{smoothing}) and~(\ref{empdist})); the empirical mean $m_i^{\star}$ is insensitive to this smoothing at the precision relevant for our analysis.

\subsection{Empirical verification of geometric laws}\label{sec:verification}

We analyse the ownership distributions of $63$ representative UTXO denominations across $72$ monthly snapshots from January 2018 to December 2023, and perform a Bayesian one-parameter inference to assess whether the geometric (bosonic) law predicted for identical digital wealth is realised in practice.

\begin{figure}[ht]
	\centering
	\includegraphics[width=0.45\textwidth]{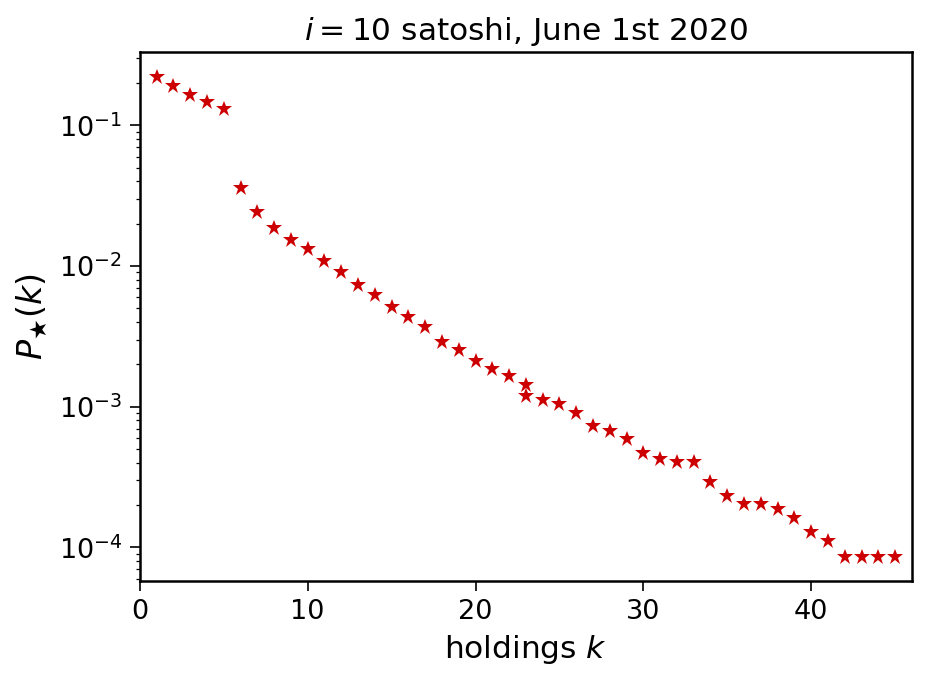}~
	\includegraphics[width=0.45\textwidth]{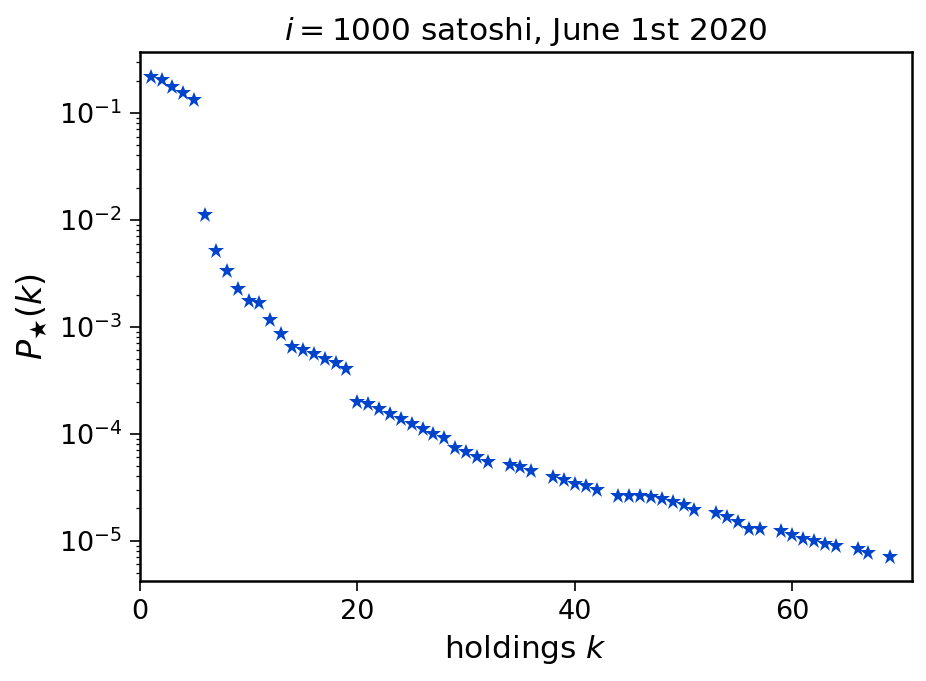}
	\caption{Log-scale empirical distributions for two UTXO denominations on June 1, 2020:
		$i = 10$ satoshi (left) and $i = 1000$ satoshi (right).
		The vertical axis shows the smoothed empirical probability $\bar{P}_{\star}(k)$ on a $\log_{10}$ scale and the horizontal axis shows $k$. The near-linear decay is visible in the raw counts as well.}
	\label{figlog}
\end{figure}

We begin by inspecting representative empirical ownership distributions on a semi-logarithmic scale (Fig.~\ref{figlog}). The near-linear decay across the full range of $k$---a feature that holds both for the raw probability $P_{\star}(k)=n_{k}^{\star}/N_{\star}$ and for its smoothed counterpart---already indicates exponential (geometric) behaviour. For the quantitative analysis that follows, we apply a centred moving average of width $9$ to the raw counts $\{n_{k}^{\star}\}$ to stabilise the sparse tail at large $k$, yielding the smoothed distribution $\bar{P}_{\star}(k)$ defined in Methods (Eqs.~(\ref{smoothing}) and~(\ref{empdist})); the overbar denotes smoothing. All Bayesian fitting, Jensen--Shannon divergences, and alternative-model comparisons below use $\bar{P}_{\star}(k)$.

Because the empirical data exclude zero holdings and have a finite maximum occupancy $k_{\max,i}$ (the common support defined in Methods), we replace the idealised geometric distribution of Eq.~(\ref{twoP})---defined for $k=0,1,2,\ldots$---with a truncated geometric form,
\begin{equation}
	P_{\beta}(k)=\left(\frac{e^{\beta}-1}{1-e^{-\beta k_{\max,i}}}\right)
	e^{-\beta k},
	\qquad
	k=1,2,\ldots,k_{\max,i},
	\label{GeoFit}
\end{equation}
which is characterised by a single parameter $\beta$; the prefactor ensures the normalisation $\sum_{k=1}^{k_{\max,i}} P_{\beta}(k)=1$. Here $\beta$ plays the role of an inverse temperature: larger $\beta$ corresponds to a ``colder'' system in which wealth is more evenly spread, while smaller $\beta$ indicates stronger concentration. For each denomination $i$ and month, the corresponding parameter $\beta_i$ is
determined by minimising the Kullback--Leibler divergence---equivalently, obtaining the maximum-likelihood estimate for $\bar{P}_{\star}(k)$ relative to $P_{\beta}(k)$---and subsequently embedded in a Bayesian framework to construct posterior distributions.

The inferred parameters yield the theoretical mean of the truncated geometric distribution,
\begin{equation}
	m=\langle k\rangle
	=\frac{1-(k_{\max,i}+1)e^{-\beta k_{\max,i}}
		+k_{\max,i}e^{-\beta(k_{\max,i}+1)}}{(1-e^{-\beta})(1-e^{-\beta k_{\max,i}})}\,,
	\label{average}
\end{equation}
which we denote by $m_i(\beta_i)$ upon substituting $\beta = \beta_i$. We first assess agreement at the level of mean holdings by comparing $m_i(\beta_i)$ of Eq.~(\ref{average}) with the empirical mean $m_i^{\star}$ of Eq.~(\ref{empiricalm}).

Beyond this mean-based comparison, we quantify the similarity between the full empirical and theoretical distributions using the Jensen--Shannon divergence $D_{\mathrm{JS}}$~\cite{Lin1991,Endres2003}, defined with a base-2 logarithm as
\begin{equation}
	D_{\mathrm{JS}}(\bar{P}_{\star},P_{\beta}) =
	\sum_{k=1}^{k_{\max,i}}~
	\frac{1}{2}\bar{P}_{\star}(k)\log_2\!\left(\frac{2\bar{P}_{\star}(k)}{\bar{P}_{\star}(k)+P_{\beta}(k)}\right)
	+\frac{1}{2}P_{\beta}(k)\log_2\!\left(\frac{2P_{\beta}(k)}{\bar{P}_{\star}(k)+P_{\beta}(k)}\right)\,.
	\label{DJS}
\end{equation}
Smaller values of $D_{\mathrm{JS}}$ indicate greater agreement, with $D_{\mathrm{JS}}\to 0$ corresponding to identical distributions. In practice, values $D_{\mathrm{JS}}\lesssim 0.1$ indicate strong similarity, as in Refs.~\cite{LIGOScientific:2018mvr,Acker2025,Yin2025}.

For each denomination $i$ and each monthly snapshot, we determine the best-fit parameter~$\beta_i$ and compute the corresponding theoretical mean $m_i(\beta_i)$, which is compared with the empirical mean~$m_i^{\star}$. Applying this procedure to all $63$ denominations over $72$ months yields the time series of $\beta_i$, $m_i$, and $m_i^{\star}$.

\begin{figure}[ht]
	\centering
	\includegraphics[width=0.45\textwidth]{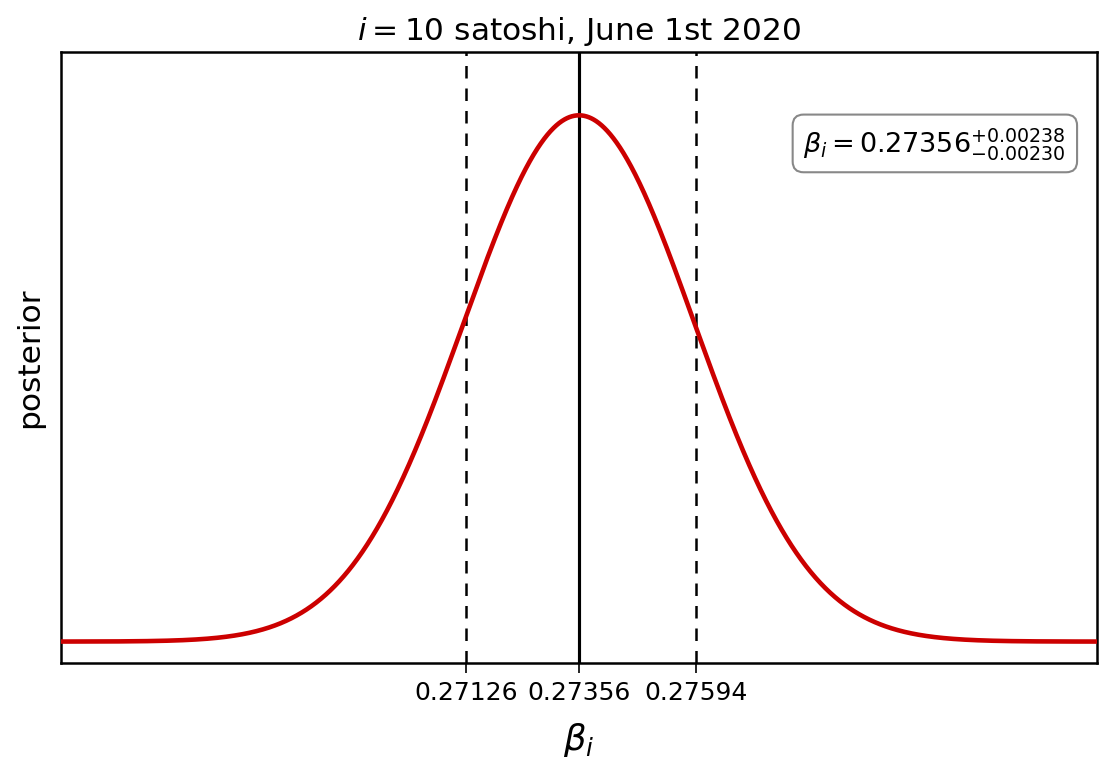}~
	\includegraphics[width=0.45\textwidth]{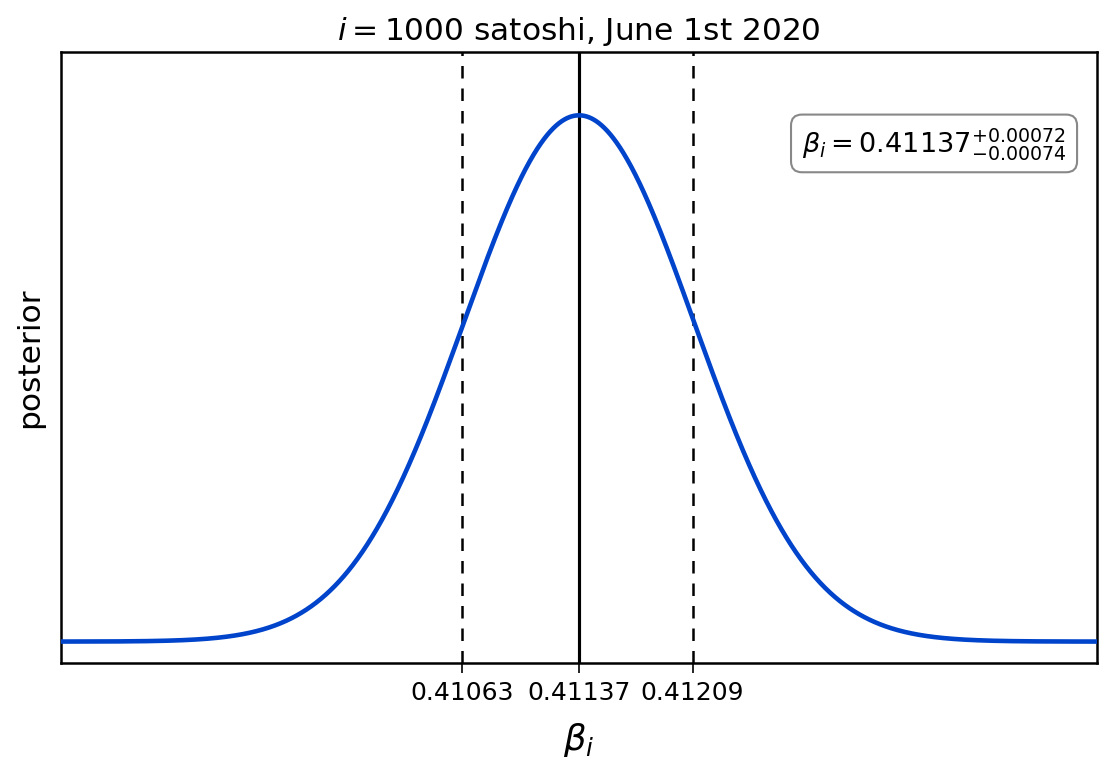}
	\caption{Bayesian posterior distributions of the inverse-temperature parameter~$\beta_i$
		for two UTXO denominations based on data from June 1, 2020:
		$i=10$ satoshi (left) and $i=1000$ satoshi (right).
		Each curve shows the normalised posterior~$p(\beta_i|\mathcal{D}_i)$,
		with vertical dashed lines indicating the 68\% credible intervals
		and solid vertical lines the maximum-posterior estimates
		$\beta_i$. Both posteriors are sharply peaked, yielding well-constrained values
		of~$\beta_i$ that quantify the geometric behaviour of each UTXO denomination.
	}
	\label{figposterior}
\end{figure}

\begin{figure}[ht]
	\centering
	\includegraphics[width=0.5\textwidth]{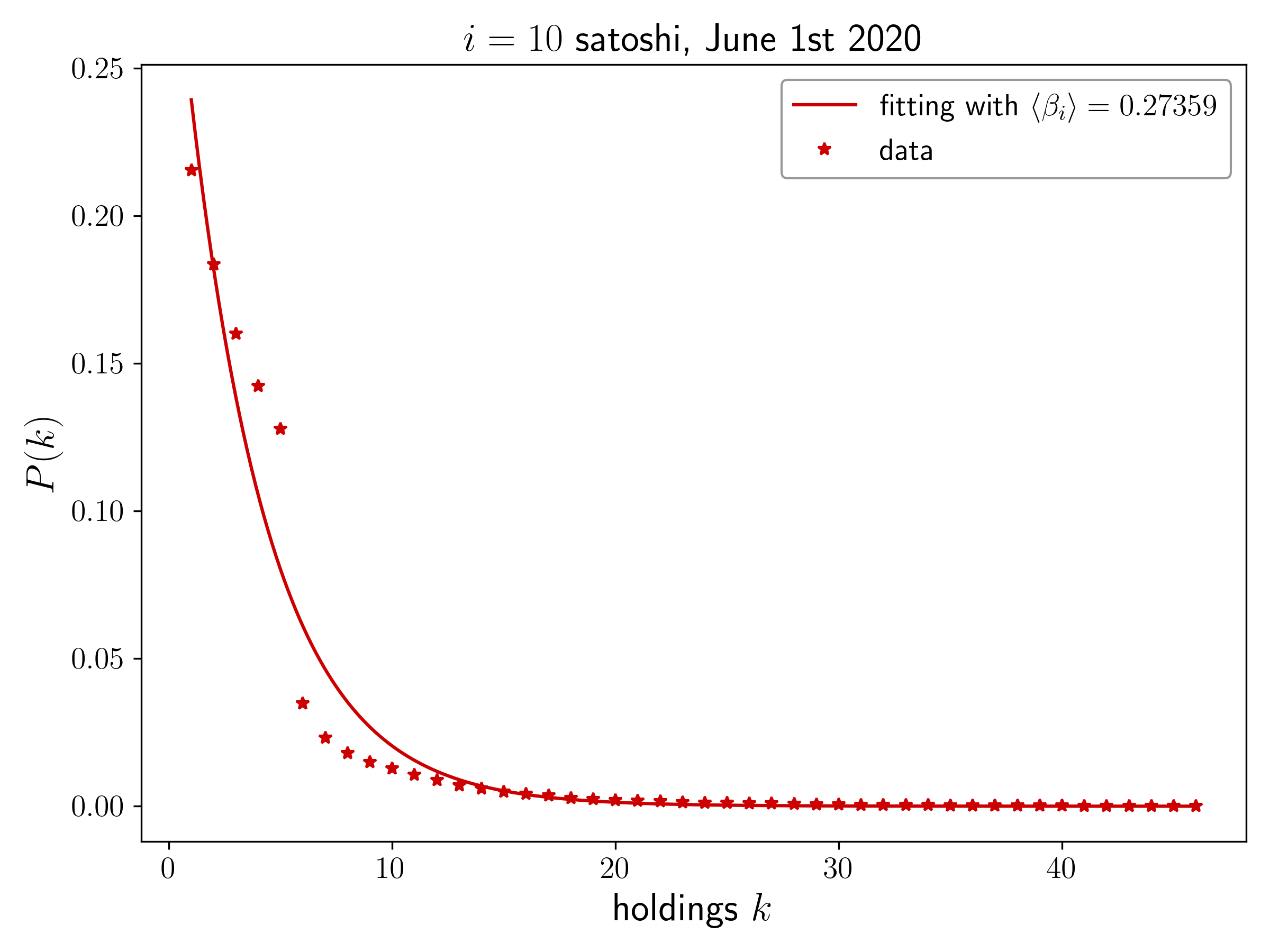}~
	\includegraphics[width=0.5\textwidth]{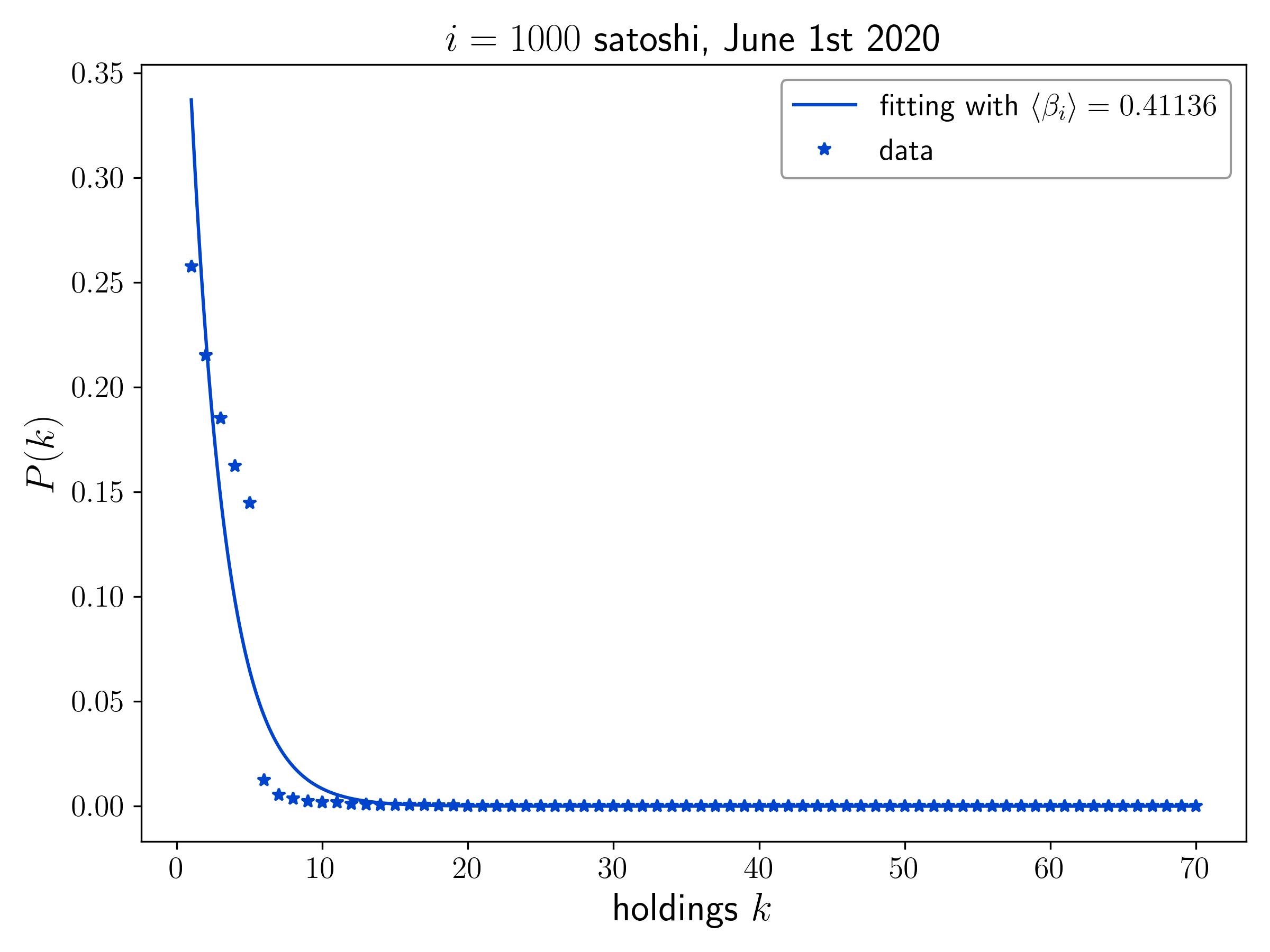}
	\caption{Geometric-distribution fits obtained using the inferred parameters~$\beta_i$
		from Fig.~\ref{figposterior}, for the same two UTXO denominations.
		The markers show the smoothed empirical probabilities~$\bar{P}_{\star}(k)$,
		and the solid lines represent the fitted geometric distributions
		$P_{\beta}(k)$ computed from Eq.~(\ref{GeoFit}).
		The fitted curves closely follow the empirical data across the full range of~$k$,
		confirming that the geometric law with the inferred~$\beta_i$
		accurately reproduces the Bitcoin UTXO ownership distributions.
	}
	\label{figfit}
\end{figure}

\begin{figure}[ht]
	\centering
	\includegraphics[width=0.5\textwidth]{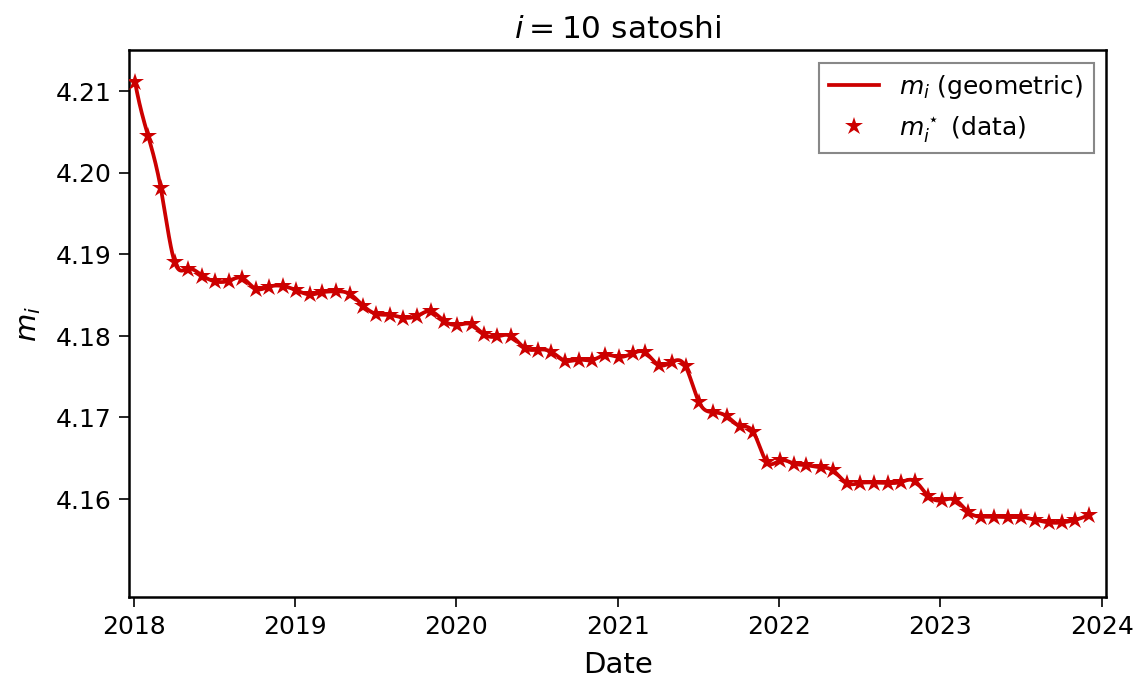}~
	\includegraphics[width=0.5\textwidth]{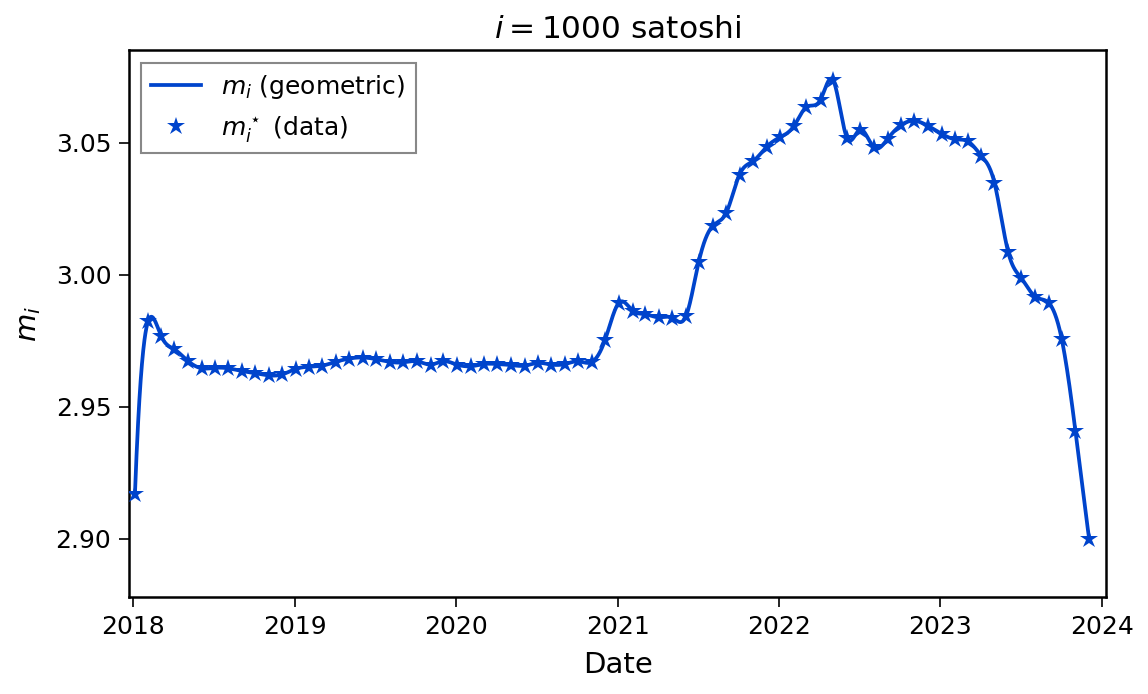}
	\caption{Monthly evolution of the empirical means $m_i^{\star}$ (stars, Eq.~(\ref{empiricalm}))
		and theoretical means $m_i$ (solid lines, Eq.~(\ref{average}))
		for the $i=10$ and $i=1000$ satoshi UTXOs from January 2018 to December 2023.
		Theoretical predictions and empirical values evolve in near-perfect synchrony, showing that the inverse-temperature parameter extracted from the data faithfully tracks the empirical mean holdings, and thus
		confirming that the geometric (bosonic) law accurately describes
		the Bitcoin UTXO ownership distributions over time.
    }
	\label{figtimeevol}
\end{figure}

Figures~\ref{figposterior}, \ref{figfit}, and \ref{figtimeevol} illustrate representative cases ($i=10$ and $i=1000$ satoshi). The posterior distributions of $\beta_i$ are sharply peaked (Fig.~\ref{figposterior}), yielding well-constrained parameter estimates. The corresponding geometric fits (Fig.~\ref{figfit}) closely reproduce the empirical distributions across the full range of $k$. The time evolution of the theoretical and empirical means (Fig.~\ref{figtimeevol}) shows near-perfect agreement over the entire observation period. Equivalent Bayesian analyses for all $63$ UTXO denominations yield the same level of agreement and are presented in the Supplementary Information (SI).

\begin{figure}[ht]
	\centering
	\includegraphics[width=0.45\textwidth]{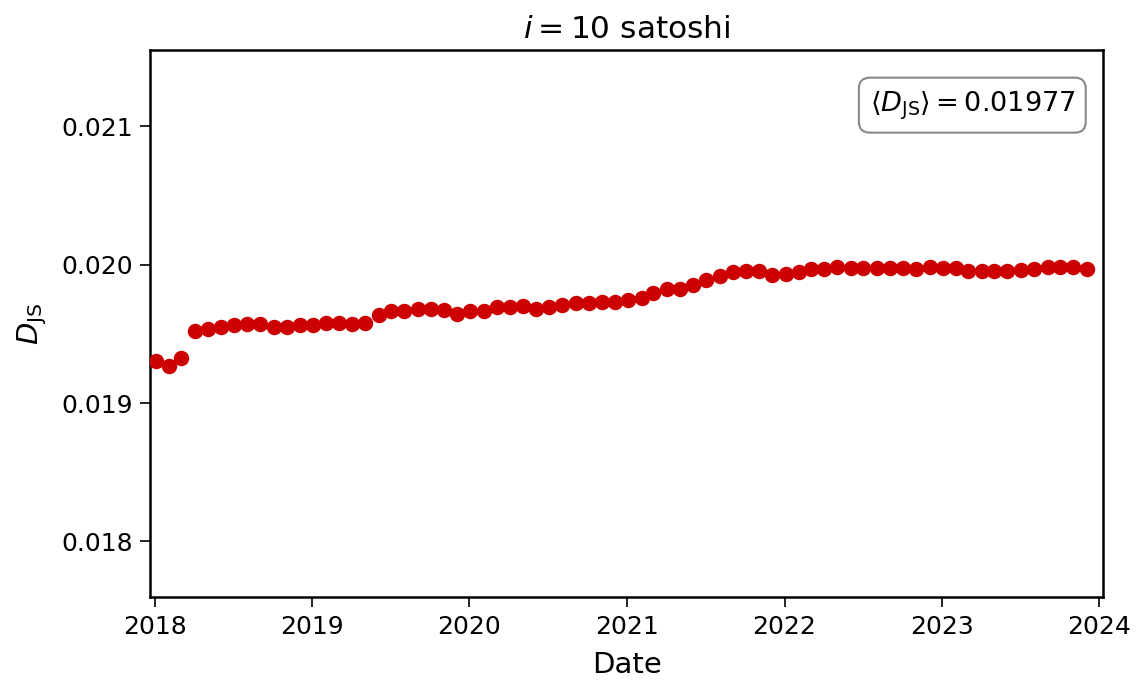}
	\includegraphics[width=0.45\textwidth]{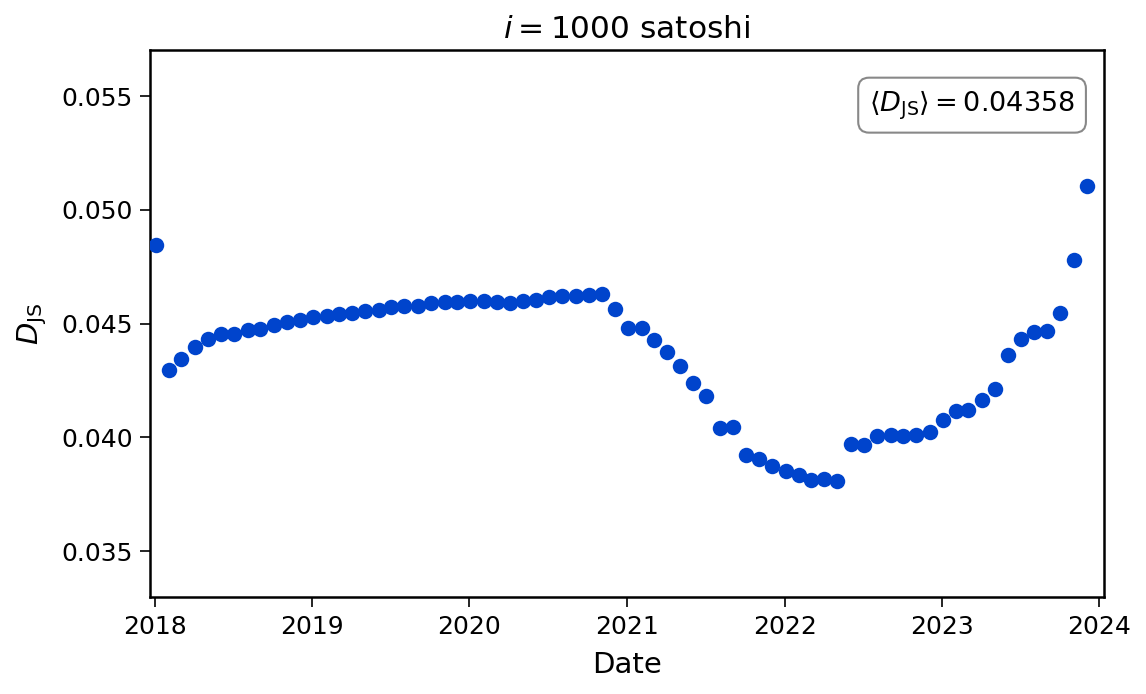}
	\caption{The Jensen--Shannon divergence, $D_{\mathrm{JS}}$, quantifying the statistical distance between the smoothed empirical and fitted distributions, $\bar{P}_{\star}(k)$ and $P_{\beta}(k)$,
		for the $i=10$ and $i=1000$ satoshi UTXOs from January 2018 to December 2023.
	}
	\label{figJSD}
\end{figure}

The Jensen--Shannon divergence further confirms the quality of the fit. As shown in Fig.~\ref{figJSD} and Table~\ref{tab:DJS_bins}, the vast majority of cases exhibit small divergence values, with a mean $\langle D_{\mathrm{JS}}\rangle=0.0448$ and $99.74\%$ of samples below $0.08$, demonstrating excellent agreement between model and data across all denominations and time snapshots.

\begin{table}[t]
	\caption{\textbf{Jensen--Shannon divergence:} $D_{\mathrm{JS}}$~(\ref{DJS}). Binned distribution
		across all $63$ UTXO denominations and $72$ monthly snapshots
		($N=4{,}536$). Nearly all cases ($99.74\%$) fall below $D_{\mathrm{JS}}=0.08$,
        while the overall mean value is $0.0448$, with a standard deviation of $0.0132$, consistent with excellent agreement between model and data.}\label{tab:DJS_bins}
	\begin{tabular}{@{}rrrr@{}}
		\toprule
		$D_{\mathrm{JS}}$ bin~range  & Count & Cumulative count & Cumulative \,\% \\
		\midrule
		\,[0.00, 0.01] & 0   & 0     & 0.00\% \\
		(0.01, 0.02] & 159 & 159   & 3.51\% \\
		(0.02, 0.03] & 668 & 827   & 18.23\% \\
		(0.03, 0.04] & 757 & 1,584 & 34.92\% \\
		(0.04, 0.05] & 1,005 & 2,589 & 57.08\% \\
		(0.05, 0.06] & 1,581 & 4,170 & 91.93\% \\
		(0.06, 0.07] & 247 & 4,417 & 97.38\% \\
		(0.07, 0.08] & 107 & 4,524 & 99.74\% \\
		(0.08, 0.09] & 5   & 4,529 & 99.85\% \\
		(0.09, 0.10] & 5   & 4,534 & 99.96\% \\
		(0.10, 0.11] & 0   & 4,534 & 99.96\% \\
		(0.11, 0.12] & 2   & 4,536 & 100.00\% \\
		\midrule
		\multicolumn{4}{c}{\textbf{Mean:}\quad $\langle D_{\mathrm{JS}}\rangle_{\mathrm{Total}}=0.0448$\qquad\quad\textbf{Std.\,Dev.:}\quad $0.0132$} \\
		\botrule
	\end{tabular}
\end{table}

\subsection{Universality of the inverse temperature across denominations and time}\label{sec:universality}

Having established the geometric form for each denomination--month pair, we now examine how the inferred inverse-temperature parameter~$\beta_i$ depends on the denomination~$i$ and on time. This is a quantitative test of the broader prediction of Ref.~\cite{KimPark2023}, in which $\beta_i$ is the per-denomination inverse-temperature parameter of the global Bitcoin equilibrium.

\begin{figure}[ht]
	\centering
	\includegraphics[width=\linewidth]{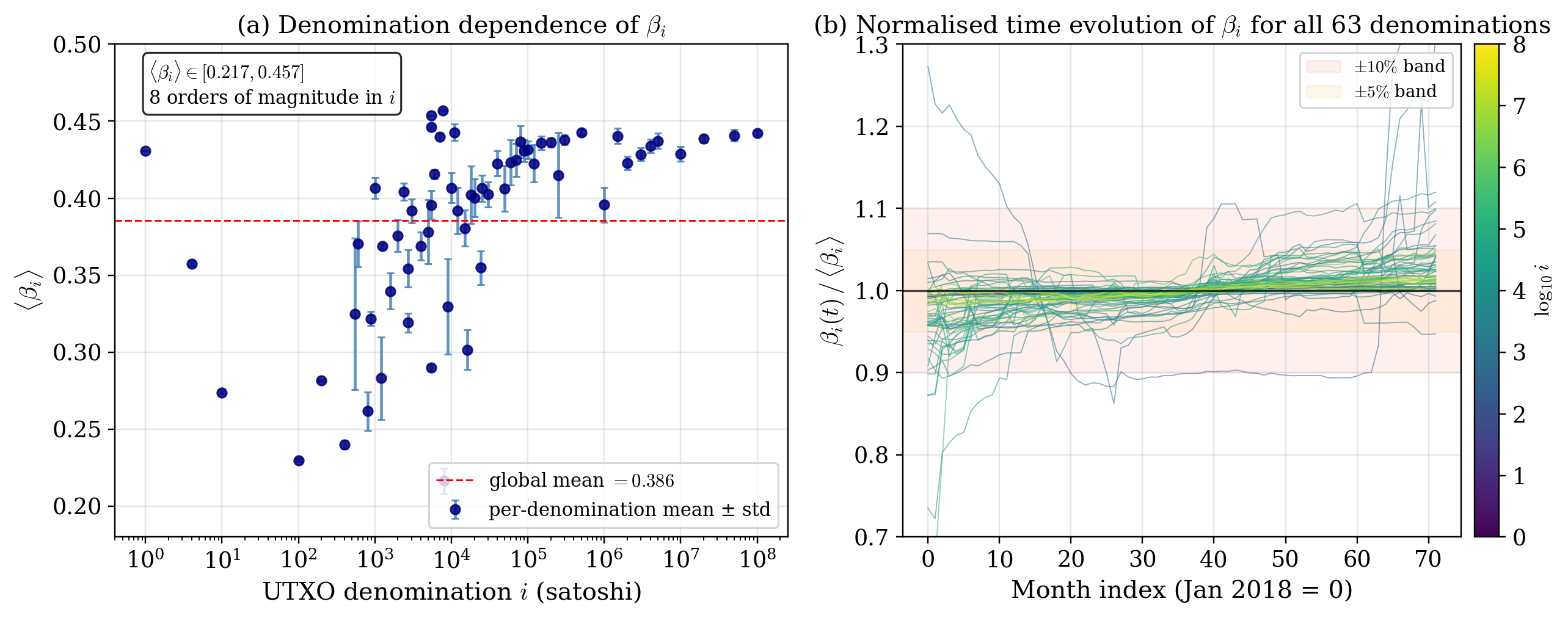}
	\caption{\textbf{Universality of the inferred inverse temperature.} (a)~Time-averaged $\langle\beta_i\rangle$ for all $63$ UTXO denominations, with error bars indicating the standard deviation over the $72$ monthly snapshots; across eight orders of magnitude in~$i$, $\langle\beta_i\rangle$ remains within a narrow band centred near $0.36$ (red dashed line). (b)~Normalised monthly trajectories $\beta_i(t)/\langle\beta_i\rangle$ for all $63$ denominations, coloured by $\log_{10}i$; the trajectories collapse onto a tight band around unity, with most denominations remaining within $\pm 5\%$ (orange shading) and nearly all within $\pm 10\%$ (red shading) over the six-year window.}
	\label{fig:beta_i}
\end{figure}

Figure~\ref{fig:beta_i}(a) shows the time-averaged $\langle\beta_i\rangle$ for each of the $63$ denominations, spanning eight orders of magnitude in $i$ (from $1$ to $10^{8}$ satoshis). The inferred values are confined to a narrow band, $\langle\beta_i\rangle\in[0.22,\,0.46]$, with a global mean of approximately $0.36$. The Pearson correlation between $\log_{10}i$ and $\langle\beta_i\rangle$ is $0.56$, indicating only a mild positive trend over eight decades; a power-law fit gives $\langle\beta_i\rangle\propto i^{0.027}$, demonstrating that the inverse temperature is approximately denomination-independent rather than strictly proportional to $i$. This near-uniformity is itself a nontrivial feature, indicating that an approximately common bosonic temperature governs holdings of very different face values---from the smallest UTXO denominations (a few satoshis) up to a full Bitcoin ($10^{8}$ satoshis).

Figure~\ref{fig:beta_i}(b) overlays the monthly evolution of $\beta_i(t)/\langle\beta_i\rangle$ for all $63$ denominations on a common scale. All trajectories collapse onto a narrow band around unity across the entire six-year window, which spans multiple Bitcoin market cycles. Quantitatively, the coefficient of variation $\sigma(\beta_i)/\langle\beta_i\rangle$ has median $1.7\%$ across the $63$ denominations, with $98.4\%$ of denominations below $10\%$ and $92.1\%$ below $5\%$ (Supplementary Information). The inverse temperature $\beta_i$ is therefore a time-invariant structural property of each denomination, not a transient feature of any one snapshot. Panels~(a) and~(b) jointly show that an approximately uniform effective bosonic temperature governs all $63$ denominations and persists over the six-year observation window.

\section{Discussion}\label{sec:discussion}

Our analysis provides strong empirical evidence that Bitcoin wealth statistics follow bosonic occupancy laws. Three mutually independent quantitative measures converge on the same conclusion across all $4{,}536$ denomination--month samples: (i)~Jensen--Shannon divergence values below $0.08$ in $99.74\%$ of cases (Table~\ref{tab:DJS_bins}); (ii)~near-perfect $\beta$--$m$ self-consistency, with the Bayesian-inferred $\beta_{i}$ matching the value obtained by inverting the analytic mean relation at Pearson correlation $r=0.99999991$ (a test that two-parameter phenomenological alternatives cannot pass by construction); and (iii)~the inverse temperature $\beta_{i}$ remaining confined to a narrow band across eight orders of magnitude in denomination, and stable to within a few percent over six years (Fig.~\ref{fig:beta_i}). These three measures collectively support the hypothesis that electronic assets behave analogously to a system of identical particles. This finding establishes a concrete bridge between statistical physics and economics, suggesting that the ontological form of wealth influences its distribution as profoundly as exchange rules or agent behaviour.

The central mechanism driving this behaviour lies in informational indistinguishability. In physical cash systems, units are distinguishable by wear, serial numbers, or spatial location, leading to Poisson-like distributions that support a broad middle class. In contrast, digital records are inherently identical; a satoshi in one address is indistinguishable from a satoshi in another. This indistinguishability reduces the combinatorial entropy of microstates associated with unit-level configurations, naturally favouring geometric distributions with heavier tails. Consequently, wealth concentration may emerge as an intrinsic structural property of the digital format itself, even in the absence of preferential attachment mechanisms, strategic accumulation, or institutional asymmetries. This offers a complementary explanation for inequality in digital economies: beyond market dynamics, part of it may also originate in the statistical mechanics of digitalisation.

While our study focuses on Bitcoin, the implications extend beyond cryptocurrencies to emerging programmable and fungible digital assets more broadly, where informational indistinguishability increasingly defines asset structure.

Several limitations of this study warrant consideration. First, our analysis defines a ``holder'' as a single Bitcoin address, without applying clustering heuristics to group addresses belonging to the same entity. While this approach preserves the raw statistical structure of the ledger, it may underestimate the concentration of wealth if large entities operate multiple addresses. However, the robustness of the geometric fit across denominations suggests that even at the address level, the underlying statistical law holds. Future work could extend this analysis to entity-level clusters to determine whether the bosonic behaviour persists at higher levels of aggregation.

Second, we excluded dust-level outputs to avoid artefacts driven by fee policies and change-making behaviour. While this focuses the analysis on economically meaningful denominations, it leaves open the question of how micro-transactions and protocol-level constraints influence the distribution tail.

Third, our theoretical framework assumes a closed system with conserved wealth, whereas the Bitcoin economy is open, with continuous entry and exit of holders and a monotonically growing total supply. The fact that the geometric law persists despite these open-system dynamics demonstrates its structural resilience, but further modelling of open bosonic systems could refine the theoretical predictions. Addressing these limitations will be important for assessing the generality of the bosonic wealth hypothesis beyond Bitcoin.

The geometric law is not introduced merely as a phenomenological fit but as a \emph{one-parameter theoretical prediction} derived from the indistinguishability postulate, in which $\beta$ and the mean $m$ are mutually determined through the analytic relation of Eq.~(\ref{average}). This mutual determinacy yields a decisive consistency test unavailable to multi-parameter phenomenological alternatives: the Bayesian-inferred $\beta_i$ and the value $\beta_i^{(m)}$ obtained by inverting Eq.~(\ref{average}) at the empirical mean $m_{i}^{\star}$ must coincide across every sample. Evaluating this over all $4{,}536$ denomination--month pairs yields Pearson correlation $r=0.99999991$, with a mean relative residual of $|\beta_{i}-\beta_i^{(m)}|/\beta_i = 4.5\times 10^{-5}$ and a maximum of $4.8\times 10^{-4}$; $89.2\%$ of samples fall below $10^{-4}$ and all samples fall below $10^{-3}$. Phenomenological models with two free parameters (log-normal, negative binomial) cannot satisfy such a test by construction, because a single empirical mean does not fix both parameters. The geometric law is thus the \emph{only} well-fitting candidate among those considered whose internal self-consistency is empirically verified.

To probe robustness further, we additionally compare the goodness of fit of the geometric against three alternatives---truncated Poisson (1~parameter), truncated negative binomial (2~parameters), and discretised log-normal (2~parameters)---via the Jensen--Shannon divergence, using a common raw-count MLE protocol (distinct from the smoothed Bayesian fit of Table~\ref{tab:DJS_bins}). Under this uniform fitting protocol, every candidate was refitted by maximum-likelihood estimation on the raw counts $\{n_{k}^{\star}\}$ over the common support, and $D_{\mathrm{JS}}$ was evaluated against the smoothed empirical distribution $\bar{P}_{\star}(k)$. As shown in Table~\ref{tab:model_comparison}, the one-parameter geometric attains the lowest mean $D_{\mathrm{JS}}=0.242$, outperforming even the two-parameter negative binomial ($0.245$) and log-normal ($0.304$); in $99.8\%$ of samples the geometric $D_{\mathrm{JS}}$ is within $0.01$ of the best-performing candidate for that sample. The Poisson distribution---the canonical model for distinguishable-unit statistics---also performs worse than the geometric (mean $D_{\mathrm{JS}}=0.251$ versus $0.242$), confirming that the data are inconsistent with distinguishable-unit statistics. Thus, adding a second free parameter yields no meaningful gain over the theoretically motivated one-parameter geometric, and the latter remains both the most parsimonious and the most physically grounded description of the data.

Beyond the divergence-based comparison, two parameter-free consistency checks reinforce the geometric interpretation. The theoretical Gini coefficient derived from the fitted $\beta_{i}$ tracks the empirical Gini coefficient with Pearson correlation $r=0.986$ across all $4{,}536$ samples, indicating that the geometric law captures the inequality structure and not just the overall shape of the distribution. A discrete Kolmogorov--Smirnov statistic yields values below $0.15$ in $97.6\%$ of samples (mean $0.099$, median $0.097$), consistent with the divergence-based thresholds. Taken collectively, the $\beta$--$m$ self-consistency, the low $D_{\mathrm{JS}}$ at all tested smoothing scales, the Gini agreement, and the Kolmogorov--Smirnov statistic constitute a set of methodologically distinct and complementary checks that all point to the same conclusion: Bitcoin UTXO ownership statistics are governed by a single-parameter bosonic geometric law.

\begin{table}[t]
	\caption{\textbf{Model comparison ($D_{\mathrm{JS}}$).} Mean Jensen--Shannon divergence between the smoothed empirical distribution $\bar{P}_{\star}(k)$ and each candidate model, averaged over all $4{,}536$ denomination--month samples. To ensure a common fitting protocol across all candidates, every model---including the geometric---was refitted by maximum-likelihood estimation on the raw counts $\{n_{k}^{\star}\}$ (this refit is distinct from the smoothed-data Bayesian fit reported in Table~\ref{tab:DJS_bins}). Lower $D_{\mathrm{JS}}$ indicates better fit; $p$ denotes the number of free parameters.}\label{tab:model_comparison}
	\begin{tabular}{@{}lrr@{}}
		\toprule
		Model & $p$ & Mean $D_{\mathrm{JS}}$ \\
		\midrule
		Geometric          & 1 & $0.242$ \\
		Negative binomial  & 2 & $0.245$ \\
		Poisson            & 1 & $0.251$ \\
		Log-normal         & 2 & $0.304$ \\
		\botrule
	\end{tabular}
\end{table}

Beyond cryptocurrencies, this perspective offers a physics-inspired lens for analysing inequality in modern economies. Traditional economic models often rely on agent-based simulations or equilibrium assumptions that prioritise behavioural factors. By contrast, the bosonic wealth hypothesis highlights the role of asset identity---or lack thereof---as a fundamental determinant of distribution. This approach could be applied to other digital contexts, such as data ownership, carbon credits, or loyalty points, where the fungibility of units may similarly drive concentration. Furthermore, the empirically observed $\langle\beta_i\rangle$ are confined to a narrow band rather than scaling proportionally with $i$ as would be expected in a single global thermal equilibrium~\cite{KimPark2023}; this near-uniformity, combined with the strong time-stability documented in Fig.~\ref{fig:beta_i}, suggests that each denomination effectively realises an approximately independent bosonic subsystem at a similar effective temperature. Interactions with other currencies, regulatory constraints, and heterogeneous usage patterns may all contribute to the residual denomination dependence and warrant further investigation.

We have presented strong empirical evidence that a real-world monetary system can exhibit bosonic occupancy statistics. The digitalisation of wealth, through informational indistinguishability, may act as a structural driver of wealth concentration, providing a candidate statistical origin for the amplification of inequality in digital economies. As the global financial system continues to evolve toward electronic and tokenised forms of value, understanding the statistical consequences of digitalisation becomes essential. The framework developed here provides a foundational step toward analysing inequality not merely as an outcome of social or institutional factors, but as an emergent property of the informational nature of wealth itself.

\section{Methods}\label{sec:methods}

\subsection*{Data collection and preprocessing}
Bitcoin mainnet data were retrieved from a fully synchronised Bitcoin Core archive node (version 0.21.0). The blockchain state was reconstructed using a custom analysis pipeline built on BlockSci~\cite{kalodner2020blocksci}. We extracted the set of unspent transaction outputs (UTXOs) at monthly intervals from January 2018 to December 2023, resulting in 72 distinct snapshots. For each snapshot, we parsed the output locking scripts to identify standard address types, including Pay-to-PubKey-Hash (P2PKH), Pay-to-Script-Hash (P2SH), and Segregated Witness (SegWit) variants. Non-standard scripts and unspendable outputs (e.g., \texttt{OP\_RETURN}) were excluded from the analysis.

A ``holder'' was defined as a unique Bitcoin address. We deliberately did not apply address-clustering heuristics (e.g., common-input-ownership) to ensure that the statistical analysis reflected the raw ledger structure rather than inferred entity ownership. UTXO denominations were selected based on frequency analysis of the entire observation period; we identified 63 face values (measured in satoshis) that were consistently overrepresented in the UTXO set. These denominations correspond to round-number payment conventions and salient value points used by wallets and services. To avoid mechanical artefacts arising from fee policies and change-making behaviour, we excluded dust outputs. An output was classified as dust if its value was below the relay fee threshold required to spend it, as defined by the Bitcoin Core policy at the time of each snapshot. Internal consistency checks were performed by comparing our extracted UTXO counts and total values against the node's reported statistics via the standard RPC interface (\texttt{gettxoutsetinfo}).

\subsection*{Statistical analysis and parameter inference}
For each denomination~$i$ and monthly snapshot, we first constructed the raw occupancy counts $\{n_{k}^{\star}\}$, where $n_{k}^{\star}$ denotes the number of addresses holding exactly $k$ UTXOs of denomination~$i$, with $1\le k\le k_{\max,i}^{\mathrm{raw}}$ (zero holdings are excluded). To stabilise the sparse tail at large $k$, we applied a centred moving average of width $w=9$,
\begin{equation}
	\bar{n}_{k}^{\star} = \frac{1}{|J_{k}|}\sum_{j\in J_{k}} n_{k+j}^{\star},
	\qquad
	J_{k} = \{j\in\mathbb{Z}: -4\le j\le 4,\ 1\le k+j\le k_{\max,i}^{\mathrm{raw}}\},
	\label{smoothing}
\end{equation}
where the average is taken only over indices that fall within the data range; in the interior ($5\le k\le k_{\max,i}^{\mathrm{raw}}-4$) the divisor $|J_{k}|=w=9$ recovers the standard centred mean of $w$ values, while at the boundaries it equals the number of available terms.
and then truncated each denomination to a common support $1\le k\le k_{\max,i}$, where $k_{\max,i}$ is the minimum of $k_{\max,i}^{\mathrm{raw}}$ over all 72 monthly snapshots, ensuring a consistent domain across time. The smoothed empirical probability distribution is denoted with an overbar and is given by
\begin{equation}
	\bar{P}_{\star}(k) = \frac{\bar{n}_{k}^{\star}}{\sum_{k^{\prime}=1}^{k_{\max,i}} \bar{n}_{k^{\prime}}^{\star}},
	\label{empdist}
\end{equation}
to be distinguished from the raw empirical probability $P_{\star}(k)=n_{k}^{\star}/N_{\star}$ introduced in Eq.~(\ref{empiricalm}). All subsequent Bayesian fitting uses this smoothed $\bar{P}_{\star}(k)$, and all Jensen--Shannon divergences reported below are evaluated against $\bar{P}_{\star}(k)$. The alternative-model comparison described in a later subsection additionally refits each candidate by maximum-likelihood estimation on the raw counts $\{n_{k}^{\star}\}$, to apply a common protocol across candidate models; $D_{\mathrm{JS}}$ is then evaluated against $\bar{P}_{\star}(k)$ as above. As an additional sensitivity check, we repeated the geometric fit (by maximum-likelihood estimation on smoothed counts, rather than via the full Bayesian pipeline) for $w\in\{1,5,9,15,21\}$: the fraction of samples with $D_{\mathrm{JS}}<0.08$ exceeds $94.7\%$ for every $w\ge 5$, while the raw, unsmoothed case ($w=1$) yields only $76.6\%$ owing to sparse-tail shot noise rather than to any change in the underlying distribution. Detailed results of this sensitivity analysis are provided in the Supplementary Information.

To test the bosonic wealth hypothesis, we fitted the empirical data to a truncated geometric distribution $P_\beta(k)$ parameterised by an inverse-temperature parameter $\beta$:
\begin{equation}
	P_{\beta}(k)=C_{\beta} \, e^{-\beta k},
\end{equation}
where $C_{\beta} = (e^{\beta}-1)/(1-e^{-\beta k_{\max,i}})$ is the normalisation constant ensuring $\sum_{k=1}^{k_{\max,i}} P_{\beta}(k)=1$. The optimal parameter $\beta_i$ for each denomination and snapshot was determined by minimising the Kullback--Leibler (KL) divergence between the smoothed empirical and theoretical distributions:
\begin{equation}
	D_{\mathrm{KL}}(\bar{P}_{\star} \| P_{\beta}) = \sum_{k=1}^{k_{\max,i}} \bar{P}_{\star}(k) \ln \!\left[ \frac{\bar{P}_{\star}(k)}{P_{\beta}(k)} \right].
\end{equation}
Here $\ln$ denotes the natural logarithm, and letting $\bar{N}_{\star}=\sum_{k=1}^{k_{\max,i}} \bar{n}_{k}^{\star}$ denote the effective sample size, the log-likelihood satisfies $\ln\mathcal{L}(\beta)=\sum_{k=1}^{k_{\max,i}} \bar{n}_{k}^{\star}\ln P_{\beta}(k)=-\bar{N}_{\star}\,D_{\mathrm{KL}}(\bar{P}_{\star}\|P_{\beta})+\mathrm{const.}$, so that minimising $D_{\mathrm{KL}}$ is equivalent to maximum-likelihood estimation. Minimisation was performed using a standard gradient-based optimisation routine. Bayesian posterior distributions for $\beta$ were constructed as $p(\beta\,|\,\mathcal{D})\propto \exp\!\bigl(-\bar{N}_{\star}\,D_{\mathrm{KL}}\bigr)$ with a uniform prior over $\beta>0$. The theoretical mean holding $m_i(\beta_i)$ was then calculated analytically from the fitted parameter using Eq.~(\ref{average}) and compared against the empirical mean $m_i^{\star}$.

To quantify the goodness-of-fit, we computed the Jensen--Shannon divergence $D_{\mathrm{JS}}$ defined in Eq.~(\ref{DJS}), which uses the base-2 logarithm so that $D_{\mathrm{JS}}\in[0,1]$. In terms of the natural-logarithm KL divergence,
\begin{equation}
	D_{\mathrm{JS}}(\bar{P}_{\star},P_{\beta}) = \frac{1}{2\ln 2}\Big[D_{\mathrm{KL}}(\bar{P}_{\star} \| P_{\mathrm{M}}) + D_{\mathrm{KL}}(P_{\beta} \| P_{\mathrm{M}})\Big],
\end{equation}
where $P_{\mathrm{M}}= \frac{1}{2}(\bar{P}_{\star} + P_{\beta})$. This symmetric measure bounds the statistical distance between $0$ (identical distributions) and $1$ (maximally divergent). We interpreted values of $D_{\mathrm{JS}} \lesssim 0.1$ as indicating strong distributional agreement, consistent with thresholds used in gravitational-wave inference, atmospheric science, and data science~\cite{LIGOScientific:2018mvr,Acker2025,Yin2025}.

\subsection*{Alternative model comparison}
To assess whether the geometric law is uniquely supported by the data, we fitted three alternative distributions to each of the $4{,}536$ denomination--month samples: truncated Poisson ($P(k)\propto \lambda^k/k!$, 1~parameter), truncated negative binomial ($P(k)\propto \binom{k+r-1}{k}p^k(1-p)^r$, 2~parameters), and discretised log-normal ($P(k)\propto k^{-1}\exp[{-(\ln k-\mu)^2/(2\sigma^2)}]$, 2~parameters). To apply a common fitting protocol across all candidates (independent of parameter count), every model---including the geometric---was refitted by maximum-likelihood estimation on the raw counts $\{n_{k}^{\star}\}$ over the common support $1\le k\le k_{\max,i}$. For each fitted model, the Jensen--Shannon divergence $D_{\mathrm{JS}}$ was then evaluated between the fitted distribution and the smoothed empirical distribution $\bar{P}_{\star}(k)$, providing a uniform measure of distributional agreement independent of the number of free parameters (Table~\ref{tab:model_comparison}). The smoothed-data Bayesian geometric fit summarised in Table~\ref{tab:DJS_bins} is unaffected by this sensitivity analysis.

\subsection*{Numerical simulations}
To validate the theoretical derivation of the ownership distributions, we performed agent-based simulations of minimal wealth exchange processes. Two distinct models were implemented, each conserving the total number of holders $N$ and total wealth units $M$.
\begin{enumerate}
	\item \textbf{Random wealth-to-receiver:} At each step, one of the $M$ labelled units was selected uniformly at random, removed from its current holder, and reassigned to a holder chosen uniformly at random from the $N$ holders. This process models the redistribution of distinguishable units.
	\item \textbf{Random sender-to-receiver:} At each step, a sender was randomly chosen from the subset of holders with $k \ge 1$ units, and one unit was transferred to a randomly chosen receiver. This process models the exchange of indistinguishable (bosonic) units.
\end{enumerate}
Simulations were run with $N=10^4$ holders and $M=5\times 10^5$ wealth units for $10^8$ steps to ensure convergence to the stationary state. The resulting stationary distributions were compared against the theoretical Poisson and geometric laws, respectively, in Figure~\ref{fig:Simulation}.

\backmatter

\bmhead{Supplementary information}

Supplementary Information contains the complete Bayesian analyses for all 63 UTXO denominations, including posterior distributions, geometric fits, and time-evolution plots.

\bmhead{Acknowledgements}

We wish to thank Hongseok Kim for helpful discussions. During the preparation of this work, the authors used Claude (Anthropic) for figure styling assistance, preliminary code drafting for robustness checks, and language proofreading. All outputs were reviewed and verified by the authors, who take full responsibility for the content of the publication.

\section*{Declarations}

\begin{itemize}
\item \textbf{Funding:} This work is supported by Basic Science Research Program through the National Research Foundation of Korea (NRF) Grants, RS-2023-NR077094 and RS-2020-NR049598 (Center for Quantum Spacetime: CQUeST).
\item \textbf{Competing interests:} The authors declare no competing interests.
\item \textbf{Author contributions:} C.J.T.\ proposed the empirical validation of the theory using Bitcoin blockchain data and contributed to the overall experimental design.
C.P.\ and J.-H.P.\ developed the theoretical framework and designed the minimal exchange simulations.
C.P.\ performed the data processing, Bayesian analysis, and numerical simulations, and prepared all figures and the Supplementary Information.
C.J.T.\ and Y.Z.\ contributed to the acquisition, validation, and interpretation of the Bitcoin UTXO data.
C.J.T.\ and J.-H.P.\ supervised the project.
All authors discussed the results and contributed to their interpretation.
All authors contributed to writing the manuscript and approved the final version.
\item \textbf{Data availability:} The Bitcoin blockchain data used in this study are publicly available via the Bitcoin network.
\item \textbf{Code availability:} The code used for data extraction (based on BlockSci), Bayesian inference, and numerical simulations is available at \url{https://github.com/Jeong24th/bosonic-wealth-bitcoin}.
\end{itemize}

\end{document}